\documentclass[amsmath, aps, preprintnumbers, showpacs, twocolumn]{revtex4}
\usepackage{amsthm,amssymb}
\usepackage{hyperref}
\usepackage{natbib}

\newcommand{\ra}{\rangle}
\newcommand{\la}{\langle}
\DeclareMathOperator{\Tr}{Tr}
\newtheorem{theorem}{Theorem}

\begin{document}
\title{A $(5,5)$ and $(6,6)$ PPT edge state}
\author{Lieven Clarisse}
\pacs{03.67.Mn}
\email{lc181@york.ac.uk}
\affiliation{Dept. of Mathematics, University of York, Heslington, York, Y010 5DD, U.K.}

\begin{abstract}
Entangled states with a positive partial transpose (PPTES) have interest both in quantum information and in the theory of positive maps. In $3\otimes 3$ there is a conjecture by Sanpera, Bru{\ss} and Lewenstein [PRA, 63, 050301] that all PPTES have Schmidt number two (or equivalently that every 2-positive map between $3\times 3$ matrices is decomposable). In order to prove or disprove the conjecture it is sufficient to look at edge PPTES. Here the rank $m$ of the PPTES and the rank $n$ of its partial transpose seem to play an important role. Until recently all known examples of edge PPTES had ranks $(4,4)$ or $(6,7)$. In a recent paper Ha and Kye [quant-ph/0509079]  managed to find edge PPTES for all ranks except $(5,5)$ and $(6,6)$. Here we complement their work and present edge PPTES with those ranks.

\end{abstract}
\maketitle

\section{Introduction}
The power of entanglement as a physical resource in quantum information and computation has motivated a wide scale study in the mathematical structure of entanglement. The breakthrough came when the Horodecki family \cite{HHH96} 
linked the question of separability to the classification of positive maps in matrix algebras. In the 1960-80s important progress was made in the classification of positive maps. A paper by Choi \cite{Choi82} in 1982 reviews the main results until then and basically contains the skeleton of present entanglement theory. 

Let $M_n$ stand for the set of all $n\times n$ complex matrices. Probably the most important result in the theory of positive maps is that every positive map between $M_2$ and $M_n$ for $n\leq3$ is decomposable. A decomposable map can be decomposed as the sum of a completely positive map (CPM) and the combination of transposition and a CPM. In entanglement theory this translates to the fact that for all states $\rho$ in $M_n\otimes M_m$, $(nm\leq 6)$ positivity of the partial transposition 
$$
(\openone\otimes T)\rho=\rho^{T_B}\geq 0
$$
 is a necessary and sufficient condition for separability. For higher dimensions this is not the case and there exist entangled states with a positive partial transposition (PPTES). From a mathematical point of view the structure of PPTES in $M_2\otimes M_4$ and $M_3\otimes M_3$ are therefore of great interest.  In the present paper we are concerned with the latter. Only a handful of examples are available in this dimension:
 \begin{enumerate}
\item The St{\o}rmer matrix \cite{Stormer82, HHH98b}
\item The Choi matrix \cite{Choi82, HL00} 
\item The $7$-parameter chessboard states \cite{BP99}. 
\item The $6$-parameter UPB states in \cite{DMSST99} and neighbourhood \cite{Pittenger02, BGR04}.
\item The Horodecki matrix \cite{Horodecki97}.
\item The Ha et.\ al.\ matrices \cite{HKP03, HK03, HK05}. In matrix structure, these matrices lie in between (A) and (B).
\end{enumerate}

Construction of PPTES is a non-trivial task, and the UPB construction is really the only known automatic procedure \cite{Alb01}. The other constructions are very much trial and error and in the spirit of P{\'o}lya's traditional mathematics professor `In order to solve this differential equation you look at it till a solution occurs to you' \cite{Polya57}. Yet, given a PPTES there are several tools available to show it is entangled: 

(i)~A first one is the so-called realignment criterion \cite{Rudolph02, CW02, CW05} which just like the partial transposition reorders matrix entries. Here entanglement is guaranteed when the trace norm of the realigned density matrix is larger than one. 

(ii)~A second option is making use of non-decomposable positive maps, or alternatively non-decomposable entanglement witness \cite{Terhal01}. This method lacks the operational character of the realignment criterion, as it is non-trivial to prove the positiveness of a map. However, we know that every PPTES can be detected by some entanglement witness and hence this criterion is a much stronger one than the realignment criterion. 

(iii)~In Ref.\ \cite{DPS03} Doherty et.\ al.\ used entanglement witnesses to devise a computational algorithm which detects all entangled states. Furthermore for a given entangled state their algorithm outputs an entanglement witnesses $W$ detecting that state. This operator $W$ can always be written in a $k$-SOS (sum of squares) form which makes it easy to prove analytically that it is indeed an entanglement witness (see the original reference for details). 

(iv) The range criterion offers a remarkable simple criterion for PPTES with small rank \cite{Horodecki97, Woronowicz76}. It dictates that for a state $\rho$ to be separable there must exist a set of product vectors $\{|a_i\ra| b_i\ra\}$ spanning the range of $\rho$ such that $\{|a_i\ra |b^*_i\ra\}$ span the range of $\rho^T_B$. In particular, we say that a state $\rho$ strongly violates the range criterion if there is no product vector $|a_i\ra| b_i\ra$ in the range of $\rho$ such that $|a_i\ra| b^*_i\ra$ is in the range of $\rho^{T_B}$. A state which strongly violates the range criterion will be called an edge state in view of the following theorem (see \cite{LS97,STV98,KCKL00,HLVC00,LKCH00,LKHC00}):

\begin{theorem}
(i) An edge state $\delta$ is a PPTES such that for all $\epsilon\geq0$ and all separable $|ab\rangle$,
$$
\delta-\epsilon|ab\ra\la ab|
$$
 is not positive or does not have a positive partial transpose. 

(ii) Every PPTES $\rho$ can be decomposed as
$$
\rho=(1-p)\rho_{\mbox{sep}}+p\delta,
$$
With $\rho_{\mbox{sep}}$ a separable state and $\delta$ an edge state.
\end{theorem}

This theorem implies that knowledge of edge states is sufficient to characterise PPTES.

In Ref.\ \cite{SBL00} a study of the Schmidt number \cite{TH00,EK00} of density matrices was made. In particular they conjectured that all states in $M_3\otimes M_3$ have Schmidt number two. It is easy to see that to prove this, it is sufficient to prove it for edge states. They presented a proof for edge states of rank 4.  Denoting the rank of $\rho$ by $N$ and the rank of $\rho^{T_B}$ by $M$, they then analysed the situation for edge states with different ranks $(N,M)$. Unfortunately at the time they did their analysis, only edge states of dimension $(4,4)$ and $(6,7)$ were known. Recently, in a very interesting paper \cite{HK05}, Ha and Kye found edge states for all ranks except $(5,5)$ and $(6,6)$. In section II and III we construct an edge state with rank $(5,5)$ and rank $(6,6)$ respectively. In the final section we show how to construct entanglement witnesses for the presented states and give evidence in favour of the original conjecture. For a generalisation of the conjecture to higher dimensional systems the reader is referred to \cite{HCL01}.

\section{A $(5,5)$ edge PPTES}
Consider the following $(5,5)$ state:
$$
\rho_{(5,5)}=\frac{1}{13}
\left[ \begin{array}{ccccccccc}
  0 & 0 & 0 & 0 & 0 & 0 & 0 & 0 & 0 \\
  0 & 2 & -1 & 0 & 0 & 0 & 0 & 0 & 1 \\
  0 & -1 & 1 & 0 & 0 & 0 & 0 & 0 & -1 \\
  0 & 0 & 0 & 3 & 0 & -1 & -1 & 0 & 0 \\
  0 & 0 & 0 & 0 & 0 & 0 & 0 & 0 & 0 \\
  0 & 0 & 0 & -1 & 0 & 1 & 1 & 0 & 0 \\
  0 & 0 & 0 & -1 & 0 & 1 & 1 & 0 & 0 \\
  0 & 0 & 0 & 0 & 0 & 0 & 0 & 2 & -2 \\
  0 & 1 & -1 & 0 & 0 & 0 & 0 & -2 & 3
\end{array} \right],
$$
and its partial transpose:
$$
\rho^{T_B}_{(5,5)}=\frac{1}{13}
\left[ \begin{array}{ccccccccc}
  0 & 0 & 0 & 0 & 0 & 0 & 0 & 0 & 0 \\
  0 & 2 & -1 & 0 & 0 & 0 & 0 & 0 & 0 \\
  0 & -1 & 1 & 0 & 0 & 0 & 0 & 1 & -1 \\
  0 & 0 & 0 & 3 & 0 & -1 & -1 & 0 & 1 \\
  0 & 0 & 0 & 0 & 0 & 0 & 0 & 0 & 0 \\
  0 & 0 & 0 & -1 & 0 & 1 & 0 & 0 & 0 \\
  0 & 0 & 0 & -1 & 0 & 0 & 1 & 0 & 0 \\
  0 & 0 & 1 & 0 & 0 & 0 & 0 & 2 & -2 \\
  0 & 0 & -1 & 1 & 0 & 0 & 0 & -2 & 3
\end{array} \right].
$$
It is not so hard to verify that both operators are positive semi-definite and have rank 5. An analytical expression of their eigenvectors and eigenvalues is however quite complex. To show that $\rho_{(5,5)}$ is an edge state we will show that it violates the strong range criterion. For this we will use the `divide and conquer technique' from \cite{Horodecki97}.

It is not so hard to see that every vector in the range of $\rho_{(5,5)}$ can be written in the form
$$
V=(0,A,-E-F,C,0,D,D,E,F), \quad A,C,D,E,F \in \mathbb{C}.
$$
Now we have look at those vectors which can be written as a product
$$
V=(s,t,v)\otimes(x,y,z)=(sx,sy,sz,tx,ty,tz,vx,vy,vz).
$$
Taking these two conditions together we can therefore characterise all product vectors in the range of $\rho_{(5,5)}$. 

From the condition $sx=0$ we can distinguish the following sub cases:

1. $x=0, s\neq 0$, we have $vx=D=0=tz$ and therefore either $t=0$ or $z=0$. Without loss of generality we can also put $s=1$.

1.1. $t=0$, as $v(y+z)=-z$ we have $v=-z/(y+z)$. When $y=-z$, then $z=0=-y=x$ and we obtain the null vector. Thus the only case that remains is
$$
V=s\left(1,0,\frac{-z}{y+z}\right)\otimes(0,y,z),
$$
with $y\neq -z$.

1.2. $z=0$, from $ty=0$ follows that $t=0$ (otherwise $y=0$) thus this case is already covered in 1.1.

2. $s=0$, from $E=-F$ follows that $vy=-vz$. From $ty=0$ follows that either $t=0$ or $y=0$.

2.1 $t=0$, since $D=0=vx$ we have that $x=0$ and $v\neq 0$, and thus $z=-y$. Thus we get
$$
V=(0,0,v)\otimes(0,y,-y)
$$

2.2 $y=0$, and thus $vy=-vz=0$ so either $z=0$ or $v=0$.

2.2.1 $z=0$, and thus $vx=tz=0$. Because $x \neq 0$ we have that $v=0$:
$$
V=(0,t,0)\otimes(x,0,0)
$$

2.2.2 $v=0$ and thus  $vx=tz=0$. Because $t \neq 0$ we have that $z=0$ and we arrive at 2.2.1.

We now do the same analysis for $\rho^{T_B}_{(5,5)}$.  It is easy to see that every vector in its range can be written as
$$
V=(0,A,B,C,0,D,E,A+2B,-A-2B+C+D+E),
$$
with $A,B,C,D,E \in \mathbb{C}$.
We need to take the intersection of these vectors with those vectors which can be written as a product
$$
V=(s,t,v)\otimes(x,y,z)=(sx,sy,sz,tx,ty,tz,vx,vy,vz).
$$

From the condition $sx=0$ we can distinguish the following sub cases:

1.  $x=0$, and since $ty=0$ we have that $y=0$ or $t=0$

1.1. $y=0$ and thus $vy=0$ and $sy=-2sz=0$. Since $z=0$ gives the null vector we get $s=0$. Now $vz=-sy-2sz+tx+tz+vx=tz$ and thus $t=v$, and thus
$$
V=(0,t,t)\otimes(0,0,z)
$$

1.2. $t=0$, and since $vz=-sy-2sz=-vy$ either $v=0$ or $z=-y$

1.2.1. $v=0$ and thus $sy=-2sz$ or since $s=0$ is trivial we get $y=-2z$ and thus
$$
V=(s,0,0)\otimes(0,-2z,z)
$$

1.2.2. $z=-y$, and thus $-vy=-sy+2sy=sy$. Since $y=-z=0$ is trivial we get $v=-s$ and
$$
V=(-s,0,s)\otimes(0,y,-y)
$$

2. $s=0$ and thus $vy=0$, we also have $ty=0$ thus either $y=0$ or $v=t=0$, but this last case gives us again the trivial vector. From $vz=-sy-2sz+tx+ty+tz+vx=tx+tz+vx$ we have $(v-t)z=(v+t)x$. What remains is
$$
V=(0,v,t)\otimes(x,0,z),
$$
with either $z=\frac{v+t}{v-t}x$ or $x=\frac{v-t}{v+t}z$ (but only one of those in the case where $v=\pm t$).

It is now straightforward to check that when $v=(s,t,v)\otimes(x,y,z)$ belongs to the range of $\rho_{(5,5)}$ that
$v'=(s,t,v)\otimes(x^*,y^*,z^*)$  does not belong to the range of $\rho^{T_B}_{(5,5)}$. This concludes the proof that $\rho_{(5,5)}$ is a $(5,5)$ edge state.

\section{A $(6,6)$ edge PPTES}
Consider the following state
$$
\rho_{(6,6)}=\frac{1}{13}\left[ \begin{array}{ccccccccc}
  1 & 0 & 0 & 0 & 0 & 0 & 0 & 0 & -1 \\
  0 & 2 & 0 & -1 & 0 & 0 & 0 & 0 & 0 \\
  0 & 0 & 1 & 0 & 0 & 0 & 1 & 0 & 0 \\
  0 & -1 & 0 & 1 & 0 & 0 & 0 & 0 & 1 \\
  0 & 0 & 0 & 0 & 1 & 0 & 1 & 0 & 0 \\
  0 & 0 & 0 & 0 & 0 & 1 & 0 & -1 & 0 \\
  0 & 0 & 1 & 0 & 1 & 0 & 2 & 0 & 0 \\
  0 & 0 & 0 & 0 & 0 & -1 & 0 & 1 & 0 \\
  -1 & 0 & 0 & 1 & 0 & 0 & 0 & 0 & 3
\end{array} \right],
$$
with partial transpose
$$
\rho^{T_B}_{(6,6)}=\frac{1}{13}
\left[ \begin{array}{ccccccccc}
  1 & 0 & 0 & 0 & -1 & 0 & 0 & 0 & 1 \\
  0 & 2 & 0 & 0 & 0 & 0 & 0 & 0 & 0 \\
  0 & 0 & 1 & 0 & 0 & 0 & -1 & 0 & 0 \\
  0 & 0 & 0 & 1 & 0 & 0 & 0 & 1 & 0 \\
  -1 & 0 & 0 & 0 & 1 & 0 & 0 & 0 & -1 \\
  0 & 0 & 0 & 0 & 0 & 1 & 1 & 0 & 0 \\
  0 & 0 & -1 & 0 & 0 & 1 & 2 & 0 & 0 \\
  0 & 0 & 0 & 1 & 0 & 0 & 0 & 1 & 0 \\
  1 & 0 & 0 & 0 & -1 & 0 & 0 & 0 & 3
\end{array} \right].
$$
It can be checked easily that both matrices are positive definite and have rank 6. Following a similar strategy as in the previous section we now proceed to show that $\rho_{(6,6)}$ is an edge state. Every vector in the range of $\rho_{(6,6)}$ can be written in the form
\begin{align}
\label{eq1}
V=(A,B,C,D,E,F,C+E,-F,B+2D-A), 
\end{align}
with $A,B,C,D,E,F \in \mathbb{C}$, whilst every vector in the range of $\rho^{T_B}_{(6,6)}$ takes the form
\begin{align}
\label{eq2}
V'=(A,B,C,D,-A,E,E-C,D,F), 
\end{align}
with $A,B,C,D,E,F \in \mathbb{C}$. The `divide and conquer' method doesn't work so well here as we have no zeros to start with. Instead we will show directly that if the product vector
$$
V=(s,t,v)\otimes(x,y,z)=(sx,sy,sz,tx,ty,tz,vx,vy,vz).
$$
belongs to the range of $\rho_{(6,6)}$, then $V'=(s,t,v)\otimes(x^*,y^*,z^*)$ doesn't belong to the range of 
$\rho^{T_B}_{(6,6)}$. Comparing equations (\ref{eq1}) and (\ref{eq2}) with their product form we get the following set of equations:
\begin{align*}
vx&=sz+ty \\
vy&=-tz \\
vz&=sy+2tx-sx\\
ty^*&=-sx^* \\
vx^*&=tz^* - sz^* \\
vy^*&=tx^*
\end{align*}
We consider the following sub case

1. $vy\neq 0$, and it follows that all parameters are different from zero. Without loss of generality we can put $x=1$ and get
\begin{align*}
v&=sz+ty \\
vy&=-tz \\
vz&=sy+2t-s\\
ty^*&=-s \\
v&=tz^* - sz^* \\
vy^*&=t
\end{align*}
We now use $t=vy^*$ and $s=-v(y^*)^2$ to eliminate $t$ and $s$, after which we can divide every equation by $v$ resulting in
\begin{align*}
1&=-(y^*)^2z+y^*y \\
y&=-y^*z \\
z&=-(y^*)^2y+2y^*+(y^*)^2\\
1&=y^*z^* +(y^*)^2z^* \\
\end{align*}
From the first and the second equation we get $yy^*=\frac{1}{2}$, while from the second and fourth equation we get
$1=y^*(z^*-y)$ or $y=yy^*(z^*-y)=\frac{1}{2}(z^*-y)$ and thus $z=3y^*$. From the third equation we get
$$
3y^*=-(y^*)^2y+2y^*+(y^*)^2
$$
or 
$$
3=-y^*y+2+y^*
$$
and thus $y^*=\frac{3}{2}$ which contradicts $yy^*=\frac{1}{2}$. 

2. $v=y=0$, in this case it is straightforward to see that $x=y=z=0$.

3. $y=0$ and $v\neq 0$. From $sx^*=0$ and $tx^*=0$ follows that $s=t=0$ or $x=0$. In the first case we get $vx=vz=0$ or $z=x=0$ by the assumption. In the second case follows from $vz=0$ that $z=0$.

4. $v=0$ and from $tz=0$ and $tx^*=0$ follows that $t=0$ or $x=z=0$

4.1 $t=0$, from $sx^*=sz^*=0$ follows that either $s=t=v=0$ or $x=z=0=y$, but $y\neq 0$ by assumption.

4.2 $x=z=0$, from $ty=sy=0$ follows that $s=t=v=0$. 

This concludes the proof.

\section{A multitude of non-decomposable maps}
In this section we show how to construct witnesses (or alternatively positive maps) detecting the presented edge states.
A first way of constructing an entanglement witness for an edge state is the construction from Ref.\ \cite{LKCH00} as a generalisation of the similar construction for UPB states \cite{Terhal01b}:

\begin{theorem}
Let $\delta$ be an edge state with $P$ and $Q$ the projector onto the kernel of $\delta$ and $\delta^{T_B}$ respectively. If
$$
W_\delta=N(P+Q^{T_B}),
$$
with $N=1/\Tr(P+Q^{T_B})$ and
$$
\epsilon=\inf_{|ab\ra} \la ab| W_\delta |ab\ra,
$$
then $W_1=W_\delta -\epsilon \openone$
is an non-decomposable entanglement witness which detects $\delta$: $\Tr(W\delta)=-\epsilon<0$.
\end{theorem}

Although this construction is canonical, in the sense that it works for every edge state, for our specific examples, it is not the only choice. It turns out that both states violate the realignment criterion and for such states there is the following theorem (see \cite{CW04}):
\begin{theorem}
Let $\rho$ be a $M_n \otimes M_n$ state such that the realigned matrix $R(\rho)$ satisfies $\|R(\rho)\|>1$. The operator
$$
W_2=\openone-R(VU^\dagger),
$$
with unitaries $U,V$ yielding the singular value decomposition $R(\rho)=UDV^\dagger$, is an entanglement witness for $\rho$.
\end{theorem}

A third way of constructing an entanglement witness for $\rho_{(5,5)}$ and $\rho_{(6,6)}$ is using the computational algorithm by Doherty et.\ al.\ \cite{DPS03}. It turns out that the second test in their hierarchy can detect both states and produces an analytical entanglement witness $W_3$.

Numerically we calculated all three witnesses $W_1, W_2$ and $W_3$ for $\rho_{(5,5)}$ and $\rho_{(6,6)}$. After normalisation we found that all witnesses are different and yield a different value on the states. In total this gives us six different non-decomposable witnesses. In view of the conjecture that all edge states in $M_3\otimes M_3$ have Schmidt number two, we checked that these witnesses are negative on some Schmidt rank two state. This provides additional evidence in favour of the conjecture. In fact, we also verified this for the witnesses $\tilde W_i=W_i - (Tr(W_i \rho) +\epsilon)\openone$, for arbitrarily small $\epsilon>0$. These witnesses $\tilde W_i$ barely detect our states and are therefore the best candidates to check the conjecture. Note, however, that this is only evidence that our states have Schmidt number two. A full proof would consist of an explicit decomposition in Schmidt rank two states. However, finding such a decomposition is in general a very hard problem (see for instance \cite{Clarisse05b}) and we did not attempt this.

\begin{acknowledgments}
I would like to thank A. Sudbery for interesting discussions and comments on this paper. This research is part of QIP IRC www.qipirc.org (GR/S82176/01) and is in addition supported by a WW Smith Scholarship.
\end{acknowledgments}

\bibliographystyle{apsrevl}
\bibliography{ent}

\end{document}